\documentclass[%
 reprint,
superscriptaddress,
 amsmath,amssymb,
 aps,
pra,
]{revtex4-1}
\usepackage{bbm}
\usepackage{graphicx}
\usepackage{dcolumn}
\usepackage{bm}
\usepackage{braket}
\usepackage{qcircuit}

\usepackage{amsmath}
\usepackage{multirow}
\usepackage{graphicx}
\usepackage{algorithm}
\usepackage{algpseudocode}
\usepackage{comment}
\usepackage[colorinlistoftodos]{todonotes}
\usepackage[colorlinks=true, allcolors=blue]{hyperref}

\begin{document}

\title{Forbidden subspaces for level-1 QAOA and IQP circuits}
\author{Michael Streif}
\affiliation{Data:Lab, Volkswagen Group, Ungererstra{\ss}e 69, 80805 M{\"u}nchen, Federal Republic of Germany,}
\affiliation{University Erlangen-N{\"u}rnberg (FAU), Institute of Theoretical Physics, Staudtstr. 7, 91058 Erlangen, Germany}
\author{Martin Leib}
\affiliation{Data:Lab, Volkswagen Group, Ungererstra{\ss}e 69, 80805 M{\"u}nchen, Federal Republic of Germany,}
\date{\today}

\begin{abstract}
We present a thorough investigation of problems that can be solved exactly with the level-1 Quantum Approximate Optimization Algorithm (QAOA). To this end we implicitly define a class of problem Hamiltonians that employed as phase separator in a level-1 QAOA circuit provide unit overlap with a target subspace spanned by a set of computational basis states. For one-dimensional target subspaces we identify instances within the implicitly defined class of Hamiltonians for which Quantum Annealing (QA) and Simulated Annealing (SA) have an exponentially small probability to find the solution. Consequently, our results define a first demarcation line between QAOA, QA and SA, and highlight the fundamental differences between an interference-based search heuristic such as QAOA and heuristics that are based on thermal and quantum fluctuations like SA and QA respectively. Moreover, for two-dimensional solution subspaces we are able to show that the depth of the QAOA circuit grows linearly with the Hamming distance between the two target states. We further show that there are no genuine solutions for target subspaces of dimension higher than $2$ and smaller than $2^n$. We also transfer these results to Instantaneous Quantum Polynomial (IQP) circuits.
\end{abstract}

\pacs{Valid PACS appear here}
\maketitle

\section{Introduction}
The seminal developments of Shor's and Grover's algorithm, that showed a provable exponential and polynomial speedup with respect to their classical counterparts respectively, sparked the decades-long run to build a quantum computer. First quantum computing devices with tens of noisy qubits, so called Noisy Intermediate-Scale Quantum (NISQ) devices have already been built \cite{barends2016digitized,dicarlo2009demonstration,debnath2016demonstration,reagor2018demonstration}. However, to outperform todays most powerful classical computers, Shor's and Grover's algorithms require a fully error-corrected device with of the order of $10^5$ qubits \cite{jones2012layered}. Finding algorithms for NISQ devices that are superior with respect to their fastest known classical counterpart is therefore the next important step to bridge the gap to fully error corrected quantum computing. To achieve this, a deep understanding of the relations and features of NISQ quantum algorithms is essential.

Hybrid quantum classical algorithms, such as parameterized quantum circuits that are optimized in a classical learning loop, are generally believed to be the strongest candidates in the NISQ era. Among these are algorithms like  the Variational Quantum Eigensolver (VQE) \cite{peruzzo2014variational} for quantum chemistry calculations, Quantum Neural Networks (QNN) \cite{farhi2018classification,grant2018hierarchical} for machine learning tasks and the QAOA algorithm \cite{farhi2014quantum}. QAOA can be used to solve combinatorial optimization problems, e.g. MaxCut \cite{wang2018quantum}, Max E3LIN2 \cite{2014arXiv1412.6062F} and for generative machine learning tasks such as sampling from Gibbs states \cite{verdon2017quantum}. Interestingly there also exist QAOA versions of Shor's number factoring algorithm \cite{anschuetz2018variational}, and Grover's problem of searching an unstructured database \cite{jiang2017near} that substantially reduce the number of gates with respect to their counterparts for fully error-corrected quantum computers. Moreover it has been shown that there is no efficient classical algorithm that can simulate sampling from the output of a QAOA circuit \cite{farhi2016quantum}.

The performance and general characteristics of a heuristic like QAOA with a classical optimization loop are a nascent, vibrant however scattered research field \cite{streif2020training, zhou2018quantum, mbeng2019quantum, bapat2018bang, willsch2020benchmarking}. We add a piece to this puzzle by investigating the class of problems that can be solved exactly, i.e. a single run of the QAOA circuit would suffice to measure one of the possible answers to the problem which we will call target states henceforth. We coined the term "deterministic QAOA" to describe this specific setting. 

There are alternative classical- and quantum algorithms for combinatorial optimization: Quantum Annealing (QA) and its classical counterpart Simulated Annealing (SA). For these algorithms, there already exists an extensive body of research separating the strengths and weaknesses of QA and SA. Classes of problems have been identified that are either tailored \cite{mandra2018deceptive,denchev2016computational} or randomly generated and post-selected \cite{katzgraber2015seeking} to show a quantum speedup of QA, on existing hardware.  
In the present work we add QAOA to this framework of comparisons. If there is only a single target state, we are able to identify a set of problems based on their spectral features which can be solved exactly with QAOA with at most polynomially growing number of gates as a function of the problem size. Among these, there are problems that cannot be solved with neither QA nor SA, which we corroborate with their overlap distribution. We further show that for these problem instances there exists an efficient classical algorithm that can find the solution. Therefore, our results provide us with a rich understanding of the nature of the algorithm and show how interference effects separate QAOA from SA and QA.

In a recent work Aram Harrow introduces an upper bound on the ability of shallow circuits to have support on outcomes which are separated in Hamming distance \cite{napp2019efficient}. To add to this finding, we show, for the case of two target states, that the depth of a level-1 QAOA circuit has to grow linearly with the Hamming distance of the target states. We further show that it is impossible to have a level-1 deterministic QAOA circuit with a number of target states bigger than $2$ and smaller than $2^n$ where $n$ is the number of qubits. 

IQP circuits are a non-universal quantum computational paradigm for which it is known that there is no efficient classical algorithm that can simulate sampling from its output \cite{bremner2011classical, bremner2017achieving}. As such, IQP circuits are one of the examples of recent proposals to show quantum supremacy together with boson sampling \cite{lund2017quantum} and random quantum circuits \cite{arute2019quantum}. Because of the similarity of level-1 QAOA and IQP circuits we are able to transfer all insights we provide in this work to IQP circuits as well.

The present article is organized as follows: First we shortly recapitulate the QAOA algorithm in section \ref{sec:qaoa}. Then we identify a set of equations that implicitly defines instances that can be solved exactly with level-1 QAOA, cf. section \ref{sec:spectralprops}. We start by considering problems with a single target state, cf. section \ref{sec:singleTarget}. Subsequently we consider the case of two target states in section \ref{sec:2targets}. In section \ref{sec:manytargets} we show that there can be no genuine solution if the number of target states is in between $2$ and $2^n$. In section \ref{sec:iqp} we introduce IQP circuits and transfer all findings for level-1 QAOA circuits. We conclude with discussion and outlook in section \ref{sec:conclusion}.

\section{The Quantum Approximate Optimization Algorithm (QAOA)}\label{sec:qaoa}
The QAOA algorithm by Farhi et al.~\cite{farhi2014quantum} is a variational wavefunction ansatz with the goal to sample from low-energy states of a Hamiltonian $H_\mathrm{P}$ which is diagonal in the computational basis. Computational states are the simultaneous eigenstates of the Pauli-z operators $\sigma_z^{(i)}$ for all qubits, $i \in \left\{1, \dots, n \right\}$. The algorithm consists of two steps: First the expectation value $E_g$ of the Hamiltonian $H_\mathrm{P}$ for the variational state,
\begin{align}
   E_g= \min_{\Vec{\beta},\Vec{\gamma}}\braket{\Psi(\Vec{\beta},\Vec{\gamma})|H_\mathrm{P}|\Psi(\Vec{\beta},\Vec{\gamma})}\,,
    \label{minimize}
\end{align}
is repeatedly evaluated with the help of the Quantum Processing Unit (QPU) while the variational parameters $\Vec{\beta}$ and $\Vec{\gamma}$ are adjusted in an outer learning loop to minimize $E_g$. When $E_g$ is sufficiently low, the variational state is repeatedly prepared and measured to produce candidates for low-energy states of the Hamiltonian $H_\mathrm{P}$. The energy of all candidates is calculated and the lowest energy state is the outcome of the QAOA algorithm.

The ansatz for the variational wavefunction $\ket{\Psi(\Vec{\beta},\Vec{\gamma})}$ is inspired by the quantum annealing protocol where a system is initialized in an easy to prepare ground state of a local mixing Hamiltonian $H_\mathrm{X}=\sum _i\sigma_x^{(i)}$, with the 
Pauli-x operators acting on qubit $i$, $\sigma_x^{(i)}$, which is then slowly transformed to the problem Hamiltonian $H_\mathrm{P}$ \cite{kadowaki1998quantum}. The QAOA variational wavefunction resembles a trotterized version of this procedure,
\begin{align}
\ket{\Psi(\Vec{\beta},\Vec{\gamma})}=\mathrm{e}^{-\mathrm{i}\beta_pH_\mathrm{X}}\mathrm{e}^{-\mathrm{i}\gamma_pH_\mathrm{P}}\dots\mathrm{e}^{-\mathrm{i}\beta_1H_\mathrm{X}}\mathrm{e}^{-\mathrm{i}\gamma_1H_\mathrm{P}}\ket{+}\,,
\label{finalstate}
\end{align}
where the starting state $\ket{+}$ is the product state of eigenstates of $\sigma_x^{(i)}$ with eigenvalue $1$, $\ket{+} = \prod_i (\ket{0}_i + \ket{1}_i)/\sqrt{2}$, which is simultaneously the superposition of all computational basis states. The number of repetitions $p$ of the fundamental block of QAOA is called its level. This means a level-1 QAOA circuit consists of a single application of a unitary generated by the  problem Hamiltonian followed by the mixing operation generated by the mixing Hamiltonian.

To solve an arbitrary combinatorial optimization problem the first step is to reformulate its cost in terms of the energy of a Ising Hamiltonian. This diagonal Hamiltonian $H_p$ should be chosen such that it is possible to infer solutions of the combinatorial optimization problem from low-energy eigenstates. This is always possible with polynomial classical computing overhead for NP-complete combinatorial optimization problems since the spin-glass itself is a NP-complete problem. There exist various known embeddings of combinatorial optimization problems onto problem Hamiltonians $H_p$ \cite{lucas2014ising}. Low energy eigenstates from $H_p$ can be sampled with QAOA which can be recomputed to solutions of the combinatorial optimization problem.

It can be shown that the QAOA algorithm is strictly superior to QA since the step-like application of the mixing and problem Hamiltonian is the optimal solution for the optimal transport problem of transforming the initial state $\ket{+}$ to any other target state \cite{yang2017optimizing}. Various types of outer learning loops have been used thus far ranging from brute force grid search \cite{farhi2014quantum} to gradient based methods \cite{guerreschi2017practical} and recently methods inspired by supervised machine learning where the parameters $\Vec{\beta}$ and $\Vec{\gamma}$ were trained on random samples of combinatorial optimization problems and afterwards kept fixed to solve instances not seen during training of the same combinatorial optimization problem \cite{crooks2018performance,2018arXiv181204170B,zhou2018quantum, streif2020training}.

\section{Spectral Conditions for Deterministic QAOA}\label{sec:spectralprops}
In the following we derive conditions for the spectrum of the problem Hamiltonian $H_\mathrm{P}$ such that a level-1 version of QAOA ($p=1$) succeeds exactly, i.e. we consider a deterministic version of QAOA where we not only strive to minimize the expectation value $E_g$, cf. Eq.~(\ref{minimize}), but search for optimal values of $\beta$ and $\gamma$ such that we find perfect overlap, 
\begin{equation}
    1 \overset{!}{=}\text{tr}\left[\sum\limits_{t=1}^{T}\ket{t}\bra{t} \ket{\Psi(\beta, \gamma)}\bra{\Psi(\beta, \gamma)}\right]\,,
\end{equation}
with $T$ target states $\{\ket{t}\}_{t=1}^T$
that can be ground states of a generic $N$-qubit Hamiltonian that is diagonal in the computational basis, $H_\mathrm{P}=\mathrm{diag}\left(\left\{E_l |\, l \in \{0,1\}^n\right\}\right)$. To find a parametrized version of the class of spectra that fulfill the above requirement we slightly reformulate the above equality to the question if there exist complex values $\alpha_t$ for $t = {1, \dots, T}$ that are normalized $\sum_t |\alpha_t|^2 = 1$ such that, 
\begin{equation}\label{eq:overlap1}
    1\overset{!}{=}|\left(\sum\limits_{t=1}^T \alpha_t \bra{t}\right) \ket{\Psi(\beta, \gamma)} |^2\,,
\end{equation}
holds.
Since $H_\mathrm{P}$ is diagonal, the overlap of the variational QAOA wavefunction with the subspace spanned by the target states can be reformulated to, 
\begin{multline}\label{eq:overlap2}
\left(\sum\limits_{t=1}^T \alpha_t \bra{t} \right)\mathrm{e}^{-\mathrm{i}\beta H_\mathrm{X}}\mathrm{e}^{-\mathrm{i}\gamma H_\mathrm{P}}\ket{+}=\\
= \sum\limits_{l \in \{0,1\}^n} \underbrace{\frac{\mathrm{e}^{-\mathrm{i}\gamma E_l}}{\sqrt{2^n}}}_{x_l^*}\underbrace{\sum\limits_{t=1}^T \alpha_t \braket{t| \mathrm{e}^{-\mathrm{i} \beta H_\mathrm{X}}| l}}_{z_l}\,.
\end{multline}
With the definition of the complex vectors $\vec{x}$ and $\vec{z}$, cf. Eq.~(\ref{eq:overlap2}), the equality cf. Eq.~(\ref{eq:overlap1}) can be seen as the scalar product of two $2^n$-dimensional vectors, $1=|\vec{x}^* \cdot \vec{z}|$. It is easy to see that $|\vec{x}|^2 = 1$ and a small calculation, 
\begin{align*}
    |\vec{z}|^2 &= \sum\limits_{l \in \{0,1\}^n} \sum\limits_{t,t'= 1}^T  \alpha_t \alpha_{t'}^* \braket{t|\mathrm{e}^{-\mathrm{i}\beta H_\mathrm{X}}|l}\braket{l|\mathrm{e}^{\mathrm{i}\beta H_\mathrm{X}}|t'} =\\
  & = \sum\limits_{t,t' = 1}^T \alpha_t\alpha_{t'}^* \braket{t | t'} = 1 
\end{align*}
reveals $|\vec{z}|^2 = 1$ as well. From the Cauchy-Schwarz inequality we therefore know that $1=|\vec{x} \cdot \vec{z}|$ only holds for, $\vec{x} = \vec{z}$, ignoring an overall phase factor.
With this we can conclude for the phases of the complex numbers $z_l$,
\begin{equation}\label{eq:condition1}
    (\gamma E_l + \textrm{arg}(z_l))\,\textrm{mod}\, 2\pi = C \qquad \forall \hspace{1mm} l \in \{0,1\}^n\,,
\end{equation}
and for their magnitudes
\begin{equation}\label{eq:condition2}
   |z_l| = \frac{1}{\sqrt{2^n}} \qquad \forall \hspace{1mm} l \in \{0,1\}^n\,.
\end{equation}
The first conditions, cf. Eq.~(\ref{eq:condition1}), are the desired conditions on the spectrum while the second conditions, cf. Eq.~(\ref{eq:condition2}) are necessary conditions for the spectrum to exist. Since in our setting $\gamma$ is a mere rescaling of the energy spectrum we will henceforth absorb $\gamma$ into the definition of the spectrum, $\gamma E_l = \epsilon_l$.

\section{Single Target State}\label{sec:singleTarget}
If there is only a single target state $t_1$ the conditions on the magnitudes of $|z_l|$, Eq.~(\ref{eq:condition2}) are fulfilled if $\beta = \frac{1}{4}\pi, \frac{3}{4}\pi, \frac{5}{4}\pi, \frac{7}{4}\pi$ and the energy eigenvalues $E_l$ have to fulfill either the condition,
\begin{equation}\label{eq:1target_condition}
    \left(\epsilon_l - \frac{1}{2}\pi\Delta(t_1,l) \right) \bmod 2 \pi = C
\end{equation}
for $\beta=\frac{1}{4}\pi$ and $\beta=\frac{5}{4}\pi$ or
\begin{equation}
    \left(\epsilon_l - \frac{3}{2}\pi\Delta(t_1,l) \right) \bmod 2 \pi = C
\end{equation}
for $\beta=\frac{3}{4}\pi$ and $\beta=\frac{7}{4}\pi$. Here, $\Delta(t,l)$ is the Hamming distance between the computational states $l$ and the target state $t$, i.e. the number of spin flips required to change the state $l$ to the state $t$.
$C$ is an arbitrary constant that reflects the fact that energy eigenvalues are defined up to an additive constant. In the following we will concentrate on the first case, Eq.~(\ref{eq:1target_condition}), since the generalization to cases for $\beta=\frac{3}{4}\pi$ and $\beta=\frac{7}{4}\pi$ is straight forward. 
\paragraph{Construction of the Hamiltonian}
To convert the energy eigenvalues $E_l$ to a quantum circuit, we reformulate them into a Ising Hamiltonian,
\begin{align}
H_\mathrm{P}=&\sum_{i_1}^n h_{i_1} \sigma_z^{(i_1)} + \sum_{i_1,i_2}^n J_{i_1i_2}\sigma_z^{(i_1)}\sigma_z^{(i_2)}\nonumber\\&+ \sum_{i_1,i_2,i_3}^n J_{i_1i_2i_3}\sigma_z^{(i_1)} \sigma_z^{(i_2)} \sigma_z^{(i_3)} + \dots
\label{gspinglass}
\end{align}
given in terms of their on-site fields ($h_i$) and up to $k$-local interactions ($J_{i_1i_2}, J_{i_1i_2i_3}, \dots,J_{i_1i_2i_3\dots i_k}$), that fulfill the requirements of the instances found above. Here $\sigma_z^{(i)}$ are Pauli-z matrices acting on qubit $i$. To implement the evolution generated by this Hamiltonian, we transform every term to a $k$-qubit gate. To fulfill the above defined conditions on the spectrum, it is necessary to group the states according to their Hamming distance with respect to the target state $\ket{t}$ we would like to find with QAOA. We construct the Ising Hamiltonian with the help of the term
\begin{equation}\label{eq:Hammingpauli}
     \sum\limits_i^n \left(1 - 2 t^{(i)}\right)\sigma_z^{(i)}= n - 2 \tilde{\Delta}(t) \,,
\end{equation}
where $\tilde{\Delta}_t$ is the Hamming distance operator defined by the eigenstates given by the computational basis states and the eigenvalues given by the Hamming distance of the respective computational basis state and target state $\ket{t}$, $\tilde{\Delta}_t \ket{l} = \Delta(t, l) \ket{l}$ and $t^{(i)}$ is the $i$-th entry in bitstring $t$. We decompose the Ising Hamiltonians for our instances into two parts,
\begin{equation}\label{eq:sg-rep}
    H_p^{\textrm{1-target}} = \frac{\pi}{4}\sum\limits_i^n \left(1 - 2 t_1^{(i)}\right) \sigma_z^{(i)} + H_{2 \pi}\,. 
\end{equation}
The first term fixes the conditions given in Eq.~(\ref{eq:condition1}) and the second term $H_{2 \pi}$ is an arbitrary Ising Hamiltonian with the sole condition that all eigenvalues are multiples of $2\pi$, which can be adjusted for any Ising Hamiltonian by rescaling of the energies. This means that we can add a "watermark"-state $\ket{t}$ to every arbitrary Ising Hamiltonian such that QAOA deterministically creates this state which can be any state computational basis state, not necessarily the ground state.

\subsection{SA/QA-hard instances}
\label{sec-hardinstances}

Among the above defined instances there are problems that are hard to solve for both QA as well as SA. Both of these methods are heuristics designed to find a state that minimizes the energy of a given Ising Hamiltonian. 

For SA one starts in a random computational basis state and performs a random walk in the configuration space with Metropolis--Hasting updates with the goal to relax to low lying minima of the potential landscape. On the way to the solution, the found energy barriers can be overcome if their height is of the order of the thermal fluctuations or smaller. When cooling down the temperature slowly, in the best case scenario, SA finds the global minimum of the energy landscape.

For QA in comparison a system is initialized in the superposition of all computational basis states and the magnitude of the quantum fluctuations are decreased until the system settles in a minimum of the potential landscape. Tunneling has been proven to be beneficial in this process \cite{denchev2016computational}. It is however known that tunneling through a barrier is exponentially suppressed as a function of the barrier width while it is proportional to the inverse of the barrier height. 

QA therefore shows advantages compared with SA for potential landscapes where minima are separated by thin and tall barriers while both heuristics fail for minima separated by tall and wide barriers \cite{katzgraber2015seeking}. We therefore identify the two requirements for instances that are hard to solve for QA and SA: First, the potential landscape should feature a large number of minima separated by wide barriers, where the relevant metric in this case is Hamming distance. Second, only one minimum should be the global minimum  with all other minima separated by an amount of energy which is considered to be large enough such that the specific non-optimal minimum cannot be considered to be an acceptable solution to the encoded problem. 

\begin{center}
\begin{figure*}[t!]
\centering
\includegraphics[width=\textwidth]{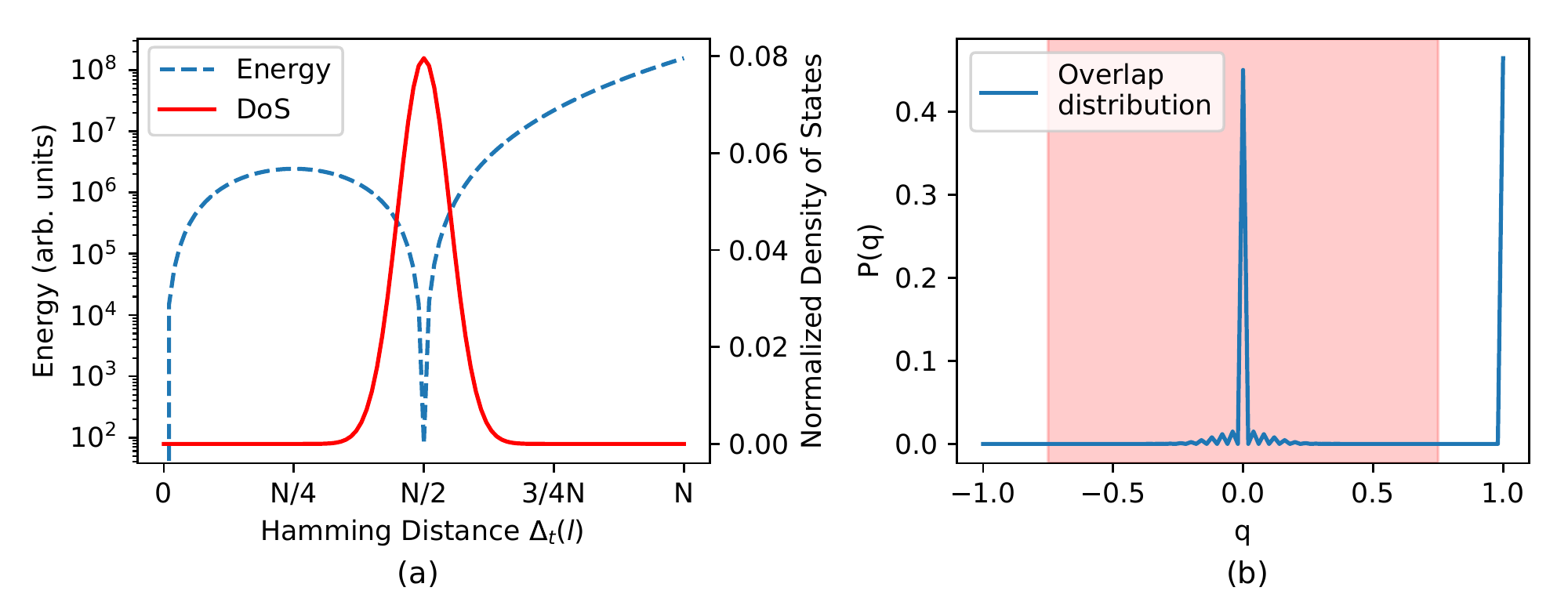}
\begin{minipage}[t!]{1\linewidth}
	\caption{(a) The dashed line shows an artificially constructed energy distribution, which fulfills the spectral conditions given in Eq.~(\ref{eq:condition1}) and employs maximal 4-local interactions. The solid line shows the normalized density of states w.r.t. the Hamming distance to highlight that the energy landscape is dominated by many sub optimal minima. (b) Overlap distribution for the given spectrum. The peak at $q=1$ denotes the overlap of every minimum with itself. The peak at in the red/dark area, however, is the overlap of all suboptimal minima with the global minimum. The position at q=0 means that they are mainly located at a Hamming distance of $n/2$. Following \cite{katzgraber2015seeking}, the peaks around $q<|0.75|$ indicate that both SA and QA will struggle to find the global minimum. Both plots show numerical data for $n=100$.}
	\label{fig:4local}
\end{minipage}
\end{figure*}
\end{center}

In general, we can generate Ising Hamiltonians with arbitrary eigenenergies. However, this could lead to $k$-local interactions up to the maximal $N$-locality. This in turn leads to a decomposition of the problem Hamiltonian block in the QAOA algorithm with an exponentially growing number of elementary gates. We therefore add an additional requirement of finite $k$-locality of the Ising Hamiltonian, where $k$ is independent of the size of the problem. The instances we found that fulfill the above requirements with maximal 4-local terms are the following, 
\begin{equation}
    H_{2 \pi}= 2\pi \tilde{\Delta}(t_1)^2 (\tilde{\Delta}(t_1) - (n/2))^2 + H_{2\pi}'\,,
\end{equation}
where $H_{2\pi}'$ is another arbitrary Ising Hamiltonian with the sole requirement that its eigenenergies are multiples of $2\pi$ and that the interactions may not be greater than 4-local. 
The quartic polynomial in the Hamming distance operator ensures that the target state is also a ground state of the Ising Hamiltonian while at the same time it generates an exponential number $\binom{n}{n/2}$ of minima with Hamming distance $n/2$. These minima are suboptimal because of the first part of Eq.~(\ref{eq:sg-rep}). In Fig.~\ref{fig:4local}~(a), we show the energy distribution as function of the Hamming distance and the density of states w.r.t. the Hamming distance. The density of states visualizes that an exponentially large fraction of random starting points in classical methods will be close to sub optimal minima.

To provide numerical evidence that these constructed instances are hard for both SA and QA and to make contact with the notions introduced in \cite{katzgraber2015seeking}, we calculate their overlap distributions. The overlap distribution is defined as the probability distribution of
\begin{align}
    q=\frac{1}{n}\sum_i^n s^{(i)}_{\alpha} s^{(i)}_{\beta}.
\end{align}
defined over two replicas, $\alpha$ and $\beta$, of the system in a thermodynamic state. It is shown that the overlap distribution allows to draw conclusions about the hardness of combinatorial problems for both Simulated Annealing and Quantum Annealing.  Instances with peaks in the overlap distribution for small values of $q$ have been identified as hard to solve for both QA and SA \cite{katzgraber2015seeking}. We calculated the overlap distributions exactly, cf. the Supplementary Material, for the 4-local instances found above where $H_{2\pi}' = 0$. We find perfect alignment of our instances with heuristics found in \cite{katzgraber2015seeking} for hard instances for QA and SA, cf. Fig.~\ref{fig:4local}~(b). 

\subsection{Classical algorithm}
The fully trained version of the QAOA circuit for instances with deterministic outcome as defined above, cf. Eq.~(\ref{eq:condition1}), does not build up any entanglement, as can be seen from the following observation,
\begin{equation}
    e^{-\mathrm{i}\left(\frac{\pi}{4}\sum\limits_i \left(1 - 2t_1^{(i)}\right)\sigma_z^{(i)} + H_{2\pi}\right)} = \prod_i e^{-\mathrm{i} \frac{\pi}{4} \left(1 - 2t_1^{(i)}\right) \sigma_z^{(i)}}\,,
\end{equation}
i.e. every gate in the fully optimized QAOA circuit is local. This suggests that there is an efficient classical algorithm to find solutions for these instances. In the following we present an efficient classical algorithm that can find the target state given oracle access to the energies of computational basis states of the Hamiltonian given in Eq.~(\ref{eq:sg-rep}): First, one queries the energy of a random computational basis state. Second, the first spin of the initial state is flipped. The Hamming distance of the resulting state w.r.t. the target state then is either increased or decreased by one. If the Hamming distance is increased  by one, then we know the initial state of the spin was the correct one. If the Hamming distance is decreased by one, then we can leave the spin as is. To see if the Hamming distance was increased or decreased we query the energy for the state with flipped spin and examine the difference to the energy of the initial state modulo $2\pi$. Since the Hamming distance can only decrease or increase by one the energy difference is either equal to $\pi/2$ if the Hamming distances increased or $-\pi/2$ if the Hamming distance decreased. We repeat the above described method for every spin and are able to find the target state with $n+1$ queries of the oracle. For a detailed description of the algorithm cf. Fig.~\ref{alg:mapping}.

\begin{figure}
\begin{algorithm}[H] 
\caption{Classical algorithm}
\begin{algorithmic}[1]
\Statex
\Function{ClassicalAlgorithm}{$H_P$}
\\
\hrulefill
\State \textbf{Input}: Oracle access to the energies of a n-spin \State \phantom{\textbf{Input}: }Hamiltonian $H_\mathrm{P}$, cf. Eq.~(\ref{eq:sg-rep})
    \State
    \textbf{Output}: ground state of $H_\mathrm{P}$\\
\hrulefill
  \State Draw random bitstring $b=(b_1,b_2,\dots,b_n)$
  \State Calculate energy $E_b=H_P(b)$
    \For{$k \gets 1$ to $n$}    
        \State{Flip spin $k \rightarrow \tilde{b}=(b_1,b_2,\dots,-b_k,\dots,b_n)$}
        \State Calculate energy $E_{\tilde{b}}=H_P(\tilde{b})$
        \If{$(E_{\tilde{b}}-E_{b}) \mod 2\pi =-\pi/2$}
          \State $b=\tilde{b}$
        \EndIf
    \EndFor
    \State \Return {b}
\EndFunction

\end{algorithmic}
\end{algorithm}
\caption{Pseudo code to find the exact solution of Eq.~(\ref{eq:sg-rep}) classically in $n+1$ queries.}
\label{alg:mapping}
\end{figure}
\section{Two target states}\label{sec:2targets}
We start the two target state case by showing under what circumstances a solution can exist. To this end we reformulate Eq.~(\ref{eq:condition2}) to, 
\begin{equation}
    | \vec{v}_l \cdot \vec{\alpha}| = \frac{1}{\sqrt{2^n}} \qquad \forall \hspace{1mm} l \in \{0,1\}^n\,,
\end{equation}
with the definition of the vector $[\vec{v}_l]_t=\bra{t} e^{-i\beta H_x} \ket{l}$. The above equation has to hold for all computational basis states $l$. Yet, it also has to hold for only 2 computational basis states. We start with the case where the Hamming distance between the two target states is odd, the case with even Hamming distance is a trivial adaption of the following. We take the two equations corresponding to the two target states, i.e. $l_1 = t_1$ and $l_2=t_2$,  and set up a linear system of equations for the unknown $\vec{\alpha}$,
\begin{multline}
    \begin{pmatrix}
    \vec{v}_{t_1}\\
    \vec{v}_{t_2}
    \end{pmatrix}
    \begin{pmatrix}
    \alpha_1\\
    \alpha_2
    \end{pmatrix} = \\
    \cos(\beta)^n
    \begin{pmatrix}
    1 & (-\mathrm{i} \tau)^{\Delta(t_1,t_2)} \\
    (-\mathrm{i}\tau)^{\Delta(t_1,t_2)} & 1
    \end{pmatrix}
     \begin{pmatrix}
    \alpha_1\\
    \alpha_2
    \end{pmatrix}=
    \begin{pmatrix}
    \frac{e^{i\varphi_1}}{\sqrt{2^n}}\\
    \frac{e^{i\varphi_2}}{\sqrt{2^n}}
    \end{pmatrix}\,,
\end{multline}
for some $\varphi_1$ and $\varphi_2$ where $\tau = \tan(\beta)$. Since the Hamming distance $\Delta(1,2)$ is odd the coefficient matrix of the above linear system of equations is non-singular and can be inverted to solve for $\vec{\alpha}$. The norm of $\vec{\alpha}$ resulting from this procedure is,
\begin{equation}
    |\alpha_1|^2 + |\alpha_2|^2 = \frac{1}{2^n\cos(\beta)^{2n}}\frac{2}{1 + \tan(\beta)^{2\Delta(t_1,t_2)}}\,,
\end{equation}
which is equal to $1$ for same values of $\beta$ as for the case with a single target state, $\beta = \frac{1}{4} \pi, \frac{3}{4} \pi, \frac{5}{4} \pi, \frac{7}{4} \pi$. If the Hamming distance between the target states is even then we take a other computational basis states such that the coefficient matrix is invertible which is always possible. Once we have found the computational basis states we can proceed through the above steps in complete analogy with the same feasible $\beta$-values as found above. We have thereby proven that only for the above cited values of $\beta$ there can be a solution and proceed in showing that there actually exists a solution by explicitly calculating it.

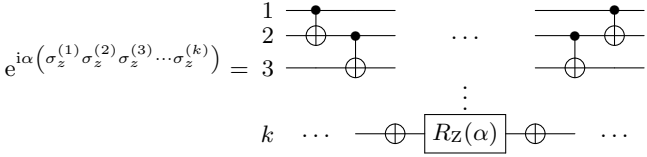
\begin{figure}[t!]
\centering
\begin{tabular}{cc}
\begin{tabular}{c}$\mathrm{e}^{\mathrm{i}\alpha\left( \sigma_z^{(1)}\sigma_z^{(2)}\sigma_z^{(3)}\cdots\sigma_z^{(k)}\right)}=$\vspace{2mm}\end{tabular}
\begin{tabular}{c}
\hspace*{0.25 cm}\Qcircuit @C=0.8em @R=.5em {
\lstick{1}& \ctrl{1} & \qw & \qw & &  & \qw  & \ctrl{1}  & \qw
\\\lstick{2}& \targ & \ctrl{1} & \qw  &\hdots &   & \ctrl{1}  & \targ  & \qw \
\\\lstick{3}& \qw & \targ & \qw &   &   & \targ  & \qw  & \qw
\\&     &        &      &     \vdots   &           &        &      &
\\&     &        & & &            &           &        &      &
\\\lstick{k} & \hdots & & \targ & \gate{R_\mathrm{Z}(\alpha)} & \targ  &\qw & \hdots  &} \end{tabular} &
\end{tabular}
\caption{Circuit realizing a $k$-local parametrized $\sigma_z$ rotation.}
\label{fig:circuit}
\end{figure}

To calculate the actual $\vec{\alpha}$ and consequently the parametrized spectrum 
we take the square of Eq.~(\ref{eq:condition2}) and use the fact that the vector $\vec{\alpha}$ is normalized,
\begin{equation}\label{eq:quadCondition}
     \vec{\alpha}^\dag \mathbf{M}_l\vec{\alpha} = 0\,
\end{equation}
where, 
\begin{equation}
    \left[\mathbf{M}_l\right]_{t,t'} = \frac{1 - \delta(t,t')}{2^n} (-i)^{\Delta(t,l) - \Delta(t',l)}\,.
\end{equation}
The above equation has to be fulfilled for all computational basis states $l$. To find out the requirements for that to happen we need to make a couple of observations first. First consider the sum and difference of Hamming distances $\Delta(t,l) \pm \Delta(t',l)$. Lets assume we approach the state l with a sequence of bit flips starting with state $t$, $t\rightarrow l_0 \rightarrow l_1 \rightarrow \dots \rightarrow l$. Every bit flip changes the sum and difference of Hamming distances by either $-2$, $0$ or $2$ starting from the Hamming distance $\Delta(t,t')$ of the two target states. Therefore an even (odd) Hamming distance between target states $t$ and $t'$ implies an even(odd) sum and difference in Hamming distances $\Delta(t,l) \pm \Delta(t',l)$. To reformulate this in a symmetric rule: Only two or no Hamming distance between three states can be odd the others are even. Second, the equality above has to hold for all $l$ therefore all contributions from off-diagonal terms have to vanish. The matrix $\mathbf{M}_l$ is hermitian and due to the special relationship of the sum and difference in Hamming distances the entries are either  $\left[\mathbf{M}_l\right]_{t,t'} = \left[\mathbf{M}_l\right]_{t',t}$ or $\left[\mathbf{M}_l\right]_{t,t'} = - \left[\mathbf{M}_l\right]_{t',t}$. This implies restrictions on the possible choices of $\alpha_t$,
\begin{align}\label{eq:hammingdist_angle}
\textrm{Re}[\alpha_1 \alpha_2^*] &= 0 \quad \textrm{if} \quad \Delta(t_1,t_2) \quad \textrm{is even} \\
\textrm{Im}[\alpha_1 \alpha_2^*] &= 0 \quad \textrm{if} \quad \Delta(t_1, t_2) \quad \textrm{is odd}\,.
\end{align}
This means if the Hamming distance between the two target states is even (odd) then the complex numbers $\alpha_1$ $\alpha_2$ are perpendicular (parallel) in the Gaussian plane. We proceed for even Hamming distance and parametrize the complex amplitudes according to $\alpha_1 = \cos(\varphi) e^{i \sigma}$ and $\alpha_2 = \pm \mathrm{i} \sin(\varphi)e^{i\sigma}$, which is the most generic parameterization that already fulfills Eq.~(\ref{eq:hammingdist_angle}) and the normalization of $\alpha$. The $z_l$ are, 
\begin{align}
z_l &= e^{i\sigma}\left(\cos(\varphi) \left(-i\right)^{\Delta(t_1,l)} \pm i \sin(\varphi)\left(-i\right)^{\Delta(t_2,l)}\right)\\
&= \exp(i(\sigma - \frac{\pi}{2}\Delta(t_1,l)\pm(-1)^{\frac{\Delta(t_2,l)-\Delta(t_1,l)}{2}} \varphi))\,.
\end{align}
The spectrum is therefore defined by,
\begin{equation}\label{eq:2targetspectrum}
    \epsilon_l = \frac{\pi}{2}\Delta(t_1,l)\pm (-1)^{\frac{\Delta(t_2,l) - \Delta(t_1,l)}{2}}\varphi\,,
\end{equation}
for arbitrary $\varphi$ and we have chosen to gauge the spectrum according to $\sigma = C$. With the definition of the energy eigenvalues we can deduce the gate sequence that needs to be executed on an all-to-all connected QPU to implement the propagation with the problem Hamiltonian. The first term on the right hand side of the conditions on the spectrum, Eq.~(\ref{eq:2targetspectrum}) can be implemented as explained above, cf. Eq.~(\ref{eq:Hammingpauli}) with local $\sigma_z$ rotations. The gate sequence for the second term can be deduced with the technique of Walsh functions \cite{Welch2014}, 
\begin{equation}
\label{eq:2targetwalsh}
    a_j = \sum\limits_{l \in \{0,1\}^n} (-1)^{\frac{\Delta(t_2,l)-\Delta(t_1,l)}{2}}(-1)^{\sum\limits_{i=1}^n j^{(i)} l^{(i)}}\,.
\end{equation}
All non-zero $a_j$ for $j\in \{0,1\}^n$ correspond to Walsh operators $\bigotimes\limits_{i=1}^n (\sigma_z^{(i)})^{j^{(i)}}$ that need to be implemented in order to get the problem Hamiltonian with the defined spectrum. Eq.~(\ref{eq:2targetwalsh}) can be reformulated to, 
\begin{equation}
    a_j = (-1)^{\sum\limits_{i=1}^n\frac{t^{(i)}_1 - t^{(i)}_2}{2}} \sum\limits_{l \in \{0,1\}^n} (-1)^{\sum\limits_{i=1}^n (t^{(i)}_2 - t^{(i)}_1 + j^{(i)})l^{(i)}}\,.
\end{equation}
From this we can deduce that there is only a single non-vanishing Walsh coefficient represented by a binary string $j$ that is one at every digit where the target states differ and zero elsewhere. This means that the second term in Eq.~(\ref{eq:2targetspectrum}) can be generated by a $\Delta(t_2,t_1)$-local term that is the tensor product of local $\sigma_z$ operators on all qubits where target state $t_1$ and target state $t_2$ differ. If we assume the generic decomposition of a k-local $\sigma_z$-rotation into two ladders of CNOTs and a local rotation Rz, cf. Fig.~\ref{fig:circuit}, this would mean that the depth of the QAOA circuit that generates a superposition of two target states scales linearly with the Hamming distance between both states.

\section{More than two target states}\label{sec:manytargets}
Based on the assumption that the only solution can be found for $\beta = \frac{1}{4} \pi, \frac{3}{4} \pi, \frac{5}{4} \pi, \frac{7}{4} \pi$ we show in the following that level-1 QAOA can not map to a genuine superposition of more than $2$ but less than $2^n$ target states. We consider a genuine superposition of target states a superposition where no amplitude vanishes. 
Eq.~(\ref{eq:condition2}) can be reformulated to 
\begin{equation}
    \cos(\beta)^n |((-i\tau)^{\Delta(t_1,l)}, \dots, (-i\tau)^{\Delta(t_T,l)} ) \vec{\alpha}| = \frac{1}{\sqrt{2^n}}\,.
\end{equation}
If there is a combination of computational basis states $l$ and $l'$ with $\Delta(t,l) = m + \Delta(t, l')$ for all $t$ and some integer $m$, it is easy to see that $|\tan(\beta)| = 1$ which is only true for $\beta = \frac{1}{4} \pi, \frac{3}{4} \pi, \frac{5}{4} \pi, \frac{7}{4} \pi$. Based on this observation as well as the calculation for the 2 target state case and numerical investigations we conducted we conjecture that the only possible solutions are $\beta = \frac{1}{4} \pi, \frac{3}{4} \pi, \frac{5}{4} \pi, \frac{7}{4} \pi$. 

Based on this conjecture what is left to show is that Eq.~(\ref{eq:quadCondition}) can only be fulfilled for all entries of $\vec{\alpha}$ but two vanishing. We start by rewriting the equation based on the above findings on Hamming distances,
\begin{equation}\label{eq:quad_lsoe}
    \sum\limits_{t>t'} (-1)^{\left \lfloor{\frac{\Delta(t,l)-\Delta(t',l)}{2}}\right \rfloor } \text{Re/Im}[\alpha_t^*\alpha_{t'}] = 0
\end{equation}
where the real and imaginary part of the product of $\vec{\alpha}$ entries is chosen according to the parity of the Hamming distance according to Eq.~(\ref{eq:Hammingdist_angle}). If we assume for the moment that all products of $\vec{\alpha}$ entries are independent, then Eq.~(\ref{eq:quad_lsoe}) can be seen as homogeneous linear system of equations where the coefficient matrix is $[(-1)^{\left \lfloor{\frac{\Delta(t,l)-\Delta(t',l)}{2}}\right \rfloor }]_{\left\{t,t'\right\},l}$. The Hamming state difference is bounded from below and above by $-\Delta(t,t')\leq \Delta(t,l) -\Delta(t', l) \leq \Delta(t,t')$, i.e. the rows in the coefficient matrix are all possible combinations of $-1$ and $1$. Therefore the coefficient matrix has full rank and the only possible solution to the linear system of equations is the trivial one,
\begin{align}\label{eq:Hammingdist_angle}
\textrm{Re}[\alpha_t \alpha_t'^*] &= 0 \quad \textrm{if} \quad \Delta(t,t') \quad \textrm{is even} \\
\textrm{Im}[\alpha_t \alpha_t'^*] &= 0 \quad \textrm{if} \quad \Delta(t, t') \quad \textrm{is odd}\,.
\end{align}
Assume a situation where there are more than two target states. Then it is always possible to choose three of them. Further assume that all three complex amplitudes $\alpha_i$ for the three target states are non-vanishing. As we already know there are two cases: either all Hamming distances between them are even or two Hamming distances are odd and the remaining one is even. If all Hamming distances are even then the real part has to vanish for all products of complex amplitudes $\alpha_t^*\alpha_{t'}$. This means that the complex vectors $\alpha_1$, $\alpha_2$ and $\alpha_3$ in the Gaussian plane must all be mutually perpendicular, which is not possible in the two-dimensional Gaussian plane. A straight forward adaption of the above reasoning also precludes the possibility of a solution in the case where two of the Hamming distances are odd.  
\section{IQP circuits}
\label{sec:iqp}
IQP circuits are a non-universal quantum computational paradigm. They serve as a tool in complexity theoretic proofs to discern quantum- from classical computational power. Efficient classical sampling from the output distribution of IQP circuits has been shown to be \#P-hard even for approximate sampling or sampling from IQP circuits with noise \cite{PhysRevLett.117.080501,bremner2017achieving}. IQP circuits are very similar to level-1 QAOA circuits: an arbitrary diagonal unitary transform $e^{-i H_p^{(\textrm{IQP})}}$, generated by the diagonal Hamiltonian $H_p^{(\textrm{IQP})}$, is applied to the equal superposition of all possible computational basis states $\ket{+}$ followed by Hadamard gates on all qubits and a measurement in the computational basis, i.e.  the IQP output state is,
\begin{equation}
    \ket{\text{IQP}}=\frac{1}{\sqrt{2}} \begin{pmatrix} 1& 1 \\ 1 & -1 \end{pmatrix}^{\otimes n} e^{-iH_p^{(\text{IQP})}}\ket{+}\,.
\end{equation}
We can transfer all findings of the preceding section, c.f. Sec.~\ref{sec:spectralprops} to IQP circuits, because of their similarity to level-1 QAOA circuits. Therefore the questions that we answer in the following is, if there are complex amplitudes $\alpha_t$ for $t=1, \dots, T$, that are normalized $\sum_t |\alpha_t|^2 = 1$ such that, 
\begin{equation}
    1 \overset{!}{=}|\left(\sum\limits_{t=1}^T \alpha_t \left\langle t \right| \right)\ket{\text{IQP}}|^2\,.
\end{equation}
Similar to the findings for QAOA we can derive from that two sets of  equations that define the spectrum,
\begin{equation}\label{eq:IQP_cond2}
    \left(\epsilon_l + \textrm{arg}\left(\sum\limits_{t=1}^T \alpha_t (-1)^{\sum\limits_{i=1}^n t^{(i)} l^{(i)}}\right)\right) \,\textrm{mod}\, 2\pi = C \,.
\end{equation}
and ensure the existence of a solution,
\begin{equation}\label{eq:IQP_cond1}
    |\sum\limits_{t=1}^T \alpha_t (-1)^{\sum\limits_{i=1}^n t^{(i)} l^{(i)}}| = 1
\end{equation}
For the single target state case the condition for the existence of a solution, cf. Eq.~(\ref{eq:IQP_cond1}), is trivially fulfilled and the spectrum is defined by, $\epsilon_l = \pi \sum_i t^{(i)} l^{(i)}$, which can be implemented by the 1-local IQP Hamiltonian,
\begin{equation}
    H_p^{(\textrm{IQP})}= \frac{\pi}{2} \sum\limits_{i=1}t_i\sigma_z^{(i)}\,,
\end{equation}
that has the same spectrum up to an overall phase factor.
In the 2-target state case the solution exists if, $Re[\alpha_1^*\alpha_2]=0$ holds. The parametrizations, $\alpha_1 = i \sin(\varphi)$ and $\alpha_2 = \cos(\varphi)$, automatically ensure the existence and normalization of the solution. With this parametrization we can deduce the spectrum to be,
\begin{equation}
    \epsilon_l = \pi \sum\limits_{i=1}^n t_1^{(i)} l^{(i)} + (-1)^{\sum\limits_{i=1}^n t_1^{(i)} l^{(i)}-\sum\limits_{i=1}^n t_2^{(i)} l^{(i)}}\varphi
\end{equation}
The first summand in the definition of the spectrum can be implemented with the Hamiltonian found in the 1-target state case we again resort to the technique of walsh functions to deduce the Hamiltonian for the second summand to get the full Hamiltonian,
\begin{equation}
    H_p^{(\textrm{IQP})}=\pi \sum\limits_{i=1}(1-t^{(1)}_i)\sigma_z^{(i)} + \varphi \bigotimes\limits_{i=1}^n \left(\sigma_z^{(i)}\right)^{\frac{1-t^{(1)}_i t^{(2)}_i}{2}}\,.
\end{equation}
In complete analogy to level-1 QAOA we find that the depth of the circuit implementing the phase seperator grows linearly with the Hamming distance between the two target states $t^{(1)}$ and $t^{(2)}$. Assuming the standard decomposition of a ladder of CNOTs and a local parametrized $R_z$-gate on an all-to-all connected QPU. 

If there are more than 2 target states we have to find solutions to the quadratic equation,
\begin{equation}\label{eq:manytarget_IQP_existence}
    \vec{\alpha}^\dag M_l \vec{\alpha} = 0\,,
\end{equation}
where, $[M_l]_{t,t'}= (-1)^{f(t,l)}(-1)^{f(t',l)}$. If we assume for a moment that all $Re[\alpha_t^* \alpha]$ are independent, we may think of Eq.~(\ref{eq:manytarget_IQP_existence}) as a linear equation for independent $Re[\alpha_t^* \alpha]$. The coefficient matrix for this system of linear equations has full rank and consequently the only solution is, $Re[\alpha_t^* \alpha]=0$ for all possible combinations $t$ and $t'$ where $t\neq t'$. This however as argued above is only possible if all $\alpha_t=0$ but two.

\section{Decay under Disorder}
In the final section we investigate the effects of disorder in the implementation of the Ising Hamiltonian $H_P$ on the unit overlap of the above introduced deterministic QAOA and IQP settings. To this end we assume a fixed but random shift acting on every eigenenergy, $\tilde{\epsilon}_l \to \epsilon_l + \delta_l$. As a generic assumption on QPU control errors we assume all $\delta_l$ to be i.i.d. with Gaussian probability distribution with mean $0$ and variance $\sigma$, i.e. $\delta_l \sim \frac{1}{\sqrt{2\pi \sigma^2}} e^{-\frac{\delta_l^2}{2\sigma^2}}$. To be able to treat QAOA and IQP circuits on common footing, we introduce the unitaries $U_{\textrm{QAOA}}$ and $U_{\textrm{IQP}}$ such that, $\ket{\Psi(\beta,\gamma)}= U_{\textrm{QAOA}}\ket{+}$ and $\ket{IQP} = U_{\textrm{IQP}}\ket{+}$. With these definitions we can derive a relation between the unitaries defined with respect to the perturbed eigenenergies $\tilde{U}_{\textrm{QAOA}}$ and $\tilde{U}_{\textrm{IQP}}$ and the respective versions with the unperturbed eigenenergies,
\begin{equation}
    \tilde{U}_{\mathrm{QAOA/IQP}} \ket{+} = \frac{1}{\sqrt{2^n}}\sum\limits_{l\in\{0,1\}^n} e^{-i\delta_l} U_{\mathrm{QAOA/IQP}} \ket{l}\,.
\end{equation}
With this we can reformulate the expectation value with respect to all possible deviations from the exact eigenenergies of the overlap of the output state of the perturbed QAOA or IQP circuit with a projector on the target subspace,
\begin{align}
    &\mathbb{E}\left[\mathrm{tr}\left[\sum\limits_{t=1}^T\ket{t}\bra{t} \tilde{U}_{\mathrm{QAOA/IQP}}\ket{+}\bra{+}\tilde{U}_{\mathrm{QAOA/IQP}}^\dag\right]\right]\nonumber\\
   &=\frac{T}{2^n} + e^{-\sigma^2}\left(1- \frac{T}{2^n}\right)\,.
\end{align}
As expected we recover the unit overlap if the disorder variance $\sigma$ vanishes. The overlap decreases exponentially with the variance of disorder making the unit overlap a volatile property. For completely random eigenenergies, $\sigma \to \infty$, the probability to end up in one of the target states is the ratio of the dimensions of the target subspace spanned by one $T=1$ or two $T=2$ computational basis states and the dimension of the entire Hilbert space $2^n$.
\section{Conclusion}
\label{sec:conclusion}
Based on the premise of deterministic QAOA, i.e. a setting in which the level-1 QAOA circuit is able to map to a genuine superposition of a set of target states, we were able to derive a number of fundamental insights. If there is a single target state the set of specially constructed optimization problems contains cases where both Quantum Annealing and Simulated Annealing fail. Consequently, our results define a first demarcation line between QAOA on one side and SA and QA on the other side. These results highlight the fundamental differences between heuristics designed to find the minimum of potential landscapes such as QA and SA and an interference-based algorithm such as QAOA where all states that are not the target state interfere destructively while only the the amplitudes of the target state add up constructively. This points to a new research direction of new encodings of combinatorial optimization problems into problem Hamiltonians $H_p$ where the desired solution is not necessarily the ground state but rather exceptional in its interference.

In the two target state case we were able to derive a parametrized family of problem Hamiltonians that generate level-1 QAOA circuits that can prepare an arbitrary superposition of two target states. With the technique of Walsh functions we were further able to show that level-1 QAOA circuit depth has to grow linearly with the Hamming distance of the target states, which contributes to ongoing research on the relation between the depth of a circuit and its computational power \cite{napp2019efficient, bravyi2018quantum}. 

Finally we argued that there is no possibility of a level-1 QAOA circuit to map to a genuine superposition of more than $2$ but less than $2^n$ target states and we transferred all of our findings to the quantum computational paradigm of IQP circuits.

\label{sec-summary}
\section*{ACKNOWLEDGMENTS}

The authors would like to thank Eddie Farhi, Jeffrey Goldstone for useful comments and enlightening discussions. We thank VW	Group	CIO	Martin	Hofmann and Director of Advanced Technologies and IT Strategy Volkswagen Group of America Florian Neukart, who	 enabled	 our	research. This project has received funding from the European Union’s Horizon 2020 research and innovation programme under the Grant Agreement No. 828826.
\bibliographystyle{unsrt}
\bibliography{1}

\begin{thebibliography}{10}

\bibitem{barends2016digitized}
Rami Barends, Alireza Shabani, Lucas Lamata, Julian Kelly, Antonio Mezzacapo,
  Urtzi Las~Heras, Ryan Babbush, Austin~G Fowler, Brooks Campbell, Yu~Chen,
  et~al.
\newblock Digitized adiabatic quantum computing with a superconducting circuit.
\newblock {\em Nature}, 534(7606):222, 2016.

\bibitem{dicarlo2009demonstration}
L~DiCarlo, JM~Chow, JM~Gambetta, Lev~S Bishop, BR~Johnson, DI~Schuster,
  J~Majer, Alexandre Blais, L~Frunzio, SM~Girvin, et~al.
\newblock Demonstration of two-qubit algorithms with a superconducting quantum
  processor.
\newblock {\em Nature}, 460(7252):240, 2009.

\bibitem{debnath2016demonstration}
Shantanu Debnath, Norbert~M Linke, Caroline Figgatt, Kevin~A Landsman, Kevin
  Wright, and Christopher Monroe.
\newblock Demonstration of a small programmable quantum computer with atomic
  qubits.
\newblock {\em Nature}, 536(7614):63, 2016.

\bibitem{reagor2018demonstration}
Matthew Reagor, Christopher~B Osborn, Nikolas Tezak, Alexa Staley, Guenevere
  Prawiroatmodjo, Michael Scheer, Nasser Alidoust, Eyob~A Sete, Nicolas Didier,
  Marcus~P da~Silva, et~al.
\newblock Demonstration of universal parametric entangling gates on a
  multi-qubit lattice.
\newblock {\em Science advances}, 4(2):eaao3603, 2018.

\bibitem{jones2012layered}
N~Cody Jones, Rodney Van~Meter, Austin~G Fowler, Peter~L McMahon, Jungsang Kim,
  Thaddeus~D Ladd, and Yoshihisa Yamamoto.
\newblock Layered architecture for quantum computing.
\newblock {\em Physical Review X}, 2(3):031007, 2012.

\bibitem{peruzzo2014variational}
Alberto Peruzzo, Jarrod McClean, Peter Shadbolt, Man-Hong Yung, Xiao-Qi Zhou,
  Peter~J Love, Al{\'a}n Aspuru-Guzik, and Jeremy~L O’brien.
\newblock A variational eigenvalue solver on a photonic quantum processor.
\newblock {\em Nature communications}, 5:4213, 2014.

\bibitem{farhi2018classification}
Edward Farhi and Hartmut Neven.
\newblock Classification with quantum neural networks on near term processors.
\newblock {\em arXiv preprint arXiv:1802.06002}, 2018.

\bibitem{grant2018hierarchical}
Edward Grant, Marcello Benedetti, Shuxiang Cao, Andrew Hallam, Joshua Lockhart,
  Vid Stojevic, Andrew~G Green, and Simone Severini.
\newblock Hierarchical quantum classifiers.
\newblock {\em arXiv preprint arXiv:1804.03680}, 2018.

\bibitem{farhi2014quantum}
Edward Farhi, Jeffrey Goldstone, and Sam Gutmann.
\newblock A quantum approximate optimization algorithm.
\newblock {\em arXiv preprint arXiv:1411.4028}, 2014.

\bibitem{wang2018quantum}
Zhihui Wang, Stuart Hadfield, Zhang Jiang, and Eleanor~G Rieffel.
\newblock Quantum approximate optimization algorithm for maxcut: A fermionic
  view.
\newblock {\em Physical Review A}, 97(2):022304, 2018.

\bibitem{2014arXiv1412.6062F}
E.~{Farhi}, J.~{Goldstone}, and S.~{Gutmann}.
\newblock A quantum approximate optimization algorithm applied to a bounded
  occurrence constraint problem.
\newblock {\em arXiv preprint arXiv:1412.6062}, December 2014.

\bibitem{verdon2017quantum}
Guillaume Verdon, Michael Broughton, and Jacob Biamonte.
\newblock A quantum algorithm to train neural networks using low-depth
  circuits.
\newblock {\em arXiv preprint arXiv:1712.05304}, 2017.

\bibitem{anschuetz2018variational}
Eric~R Anschuetz, Jonathan~P Olson, Al{\'a}n Aspuru-Guzik, and Yudong Cao.
\newblock Variational quantum factoring.
\newblock {\em arXiv preprint arXiv:1808.08927}, 2018.

\bibitem{jiang2017near}
Zhang Jiang, Eleanor~G Rieffel, and Zhihui Wang.
\newblock Near-optimal quantum circuit for grover's unstructured search using a
  transverse field.
\newblock {\em Physical Review A}, 95(6):062317, 2017.

\bibitem{farhi2016quantum}
Edward Farhi and Aram~W. Harrow.
\newblock Quantum supremacy through the quantum approximate optimization
  algorithm.
\newblock {\em arXiv preprint arXiv:1602.07674}, 2016.

\bibitem{streif2020training}
Michael Streif and Martin Leib.
\newblock Training the quantum approximate optimization algorithm without
  access to a quantum processing unit.
\newblock {\em Quantum Science and Technology}, 5(3):034008, 2020.

\bibitem{zhou2018quantum}
Leo Zhou, Sheng-Tao Wang, Soonwon Choi, Hannes Pichler, and Mikhail~D Lukin.
\newblock Quantum approximate optimization algorithm: Performance, mechanism,
  and implementation on near-term devices.
\newblock {\em arXiv preprint arXiv:1812.01041}, 2018.

\bibitem{mbeng2019quantum}
Glen~Bigan Mbeng, Rosario Fazio, and Giuseppe Santoro.
\newblock Quantum annealing: a journey through digitalization, control, and
  hybrid quantum variational schemes.
\newblock {\em arXiv preprint arXiv:1906.08948}, 2019.

\bibitem{bapat2018bang}
Aniruddha Bapat and Stephen Jordan.
\newblock Bang-bang control as a design principle for classical and quantum
  optimization algorithms.
\newblock {\em arXiv preprint arXiv:1812.02746}, 2018.

\bibitem{willsch2020benchmarking}
Madita Willsch, Dennis Willsch, Fengping Jin, Hans De~Raedt, and Kristel
  Michielsen.
\newblock Benchmarking the quantum approximate optimization algorithm.
\newblock {\em Quantum Information Processing}, 19:197, 2020.

\bibitem{mandra2018deceptive}
Salvatore Mandra and Helmut~G Katzgraber.
\newblock A deceptive step towards quantum speedup detection.
\newblock {\em Quantum Science and Technology}, 2018.

\bibitem{denchev2016computational}
Vasil~S Denchev, Sergio Boixo, Sergei~V Isakov, Nan Ding, Ryan Babbush, Vadim
  Smelyanskiy, John Martinis, and Hartmut Neven.
\newblock What is the computational value of finite-range tunneling?
\newblock {\em Physical Review X}, 6(3):031015, 2016.

\bibitem{katzgraber2015seeking}
Helmut~G Katzgraber, Firas Hamze, Zheng Zhu, Andrew~J Ochoa, and H~Munoz-Bauza.
\newblock Seeking quantum speedup through spin glasses: The good, the bad, and
  the ugly.
\newblock {\em Physical Review X}, 5(3):031026, 2015.

\bibitem{napp2019efficient}
John Napp, Rolando~L La~Placa, Alexander~M Dalzell, Fernando~GSL Brandao, and
  Aram~W Harrow.
\newblock Efficient classical simulation of random shallow 2d quantum circuits.
\newblock {\em arXiv preprint arXiv:2001.00021}, 2019.

\bibitem{bremner2011classical}
Michael~J Bremner, Richard Jozsa, and Dan~J Shepherd.
\newblock Classical simulation of commuting quantum computations implies
  collapse of the polynomial hierarchy.
\newblock {\em Proceedings of the Royal Society A: Mathematical, Physical and
  Engineering Sciences}, 467(2126):459--472, 2011.

\bibitem{bremner2017achieving}
Michael~J Bremner, Ashley Montanaro, and Dan~J Shepherd.
\newblock Achieving quantum supremacy with sparse and noisy commuting quantum
  computations.
\newblock {\em Quantum}, 1:8, 2017.

\bibitem{lund2017quantum}
AP~Lund, Michael~J Bremner, and TC~Ralph.
\newblock Quantum sampling problems, bosonsampling and quantum supremacy.
\newblock {\em npj Quantum Information}, 3(1):1--8, 2017.

\bibitem{arute2019quantum}
Frank Arute, Kunal Arya, Ryan Babbush, Dave Bacon, Joseph~C Bardin, Rami
  Barends, Rupak Biswas, Sergio Boixo, Fernando~GSL Brandao, David~A Buell,
  et~al.
\newblock Quantum supremacy using a programmable superconducting processor.
\newblock {\em Nature}, 574(7779):505--510, 2019.

\bibitem{kadowaki1998quantum}
Tadashi Kadowaki and Hidetoshi Nishimori.
\newblock {Quantum annealing in the transverse Ising model}.
\newblock {\em Physical Review E}, 58(5):5355, 1998.

\bibitem{lucas2014ising}
Andrew Lucas.
\newblock {Ising formulations of many NP problems}.
\newblock {\em Frontiers in Physics}, 2:5, 2014.

\bibitem{yang2017optimizing}
Zhi-Cheng Yang, Armin Rahmani, Alireza Shabani, Hartmut Neven, and Claudio
  Chamon.
\newblock Optimizing variational quantum algorithms using pontryagin’s
  minimum principle.
\newblock {\em Physical Review X}, 7(2):021027, 2017.

\bibitem{guerreschi2017practical}
Gian~Giacomo Guerreschi and Mikhail Smelyanskiy.
\newblock Practical optimization for hybrid quantum-classical algorithms.
\newblock {\em arXiv preprint arXiv:1701.01450}, 2017.

\bibitem{crooks2018performance}
Gavin~E Crooks.
\newblock Performance of the quantum approximate optimization algorithm on the
  maximum cut problem.
\newblock {\em arXiv preprint arXiv:1811.08419}, 2018.

\bibitem{2018arXiv181204170B}
Fernando G.~S.~L. {Brandao}, Michael {Broughton}, Edward {Farhi}, Sam
  {Gutmann}, and Hartmut {Neven}.
\newblock For fixed control parameters the quantum approximate optimization
  algorithm's objective function value concentrates for typical instances.
\newblock {\em arXiv preprint arXiv:1812.04170}, 2018.

\bibitem{Welch2014}
Jonathan Welch, Daniel Greenbaum, Sarah Mostame, and Alan Aspuru-Guzik.
\newblock {Efficient quantum circuits for diagonal unitaries without ancillas}.
\newblock {\em New Journal of Physics}, 16(3):033040, mar 2014.

\bibitem{PhysRevLett.117.080501}
Michael~J. Bremner, Ashley Montanaro, and Dan~J. Shepherd.
\newblock Average-case complexity versus approximate simulation of commuting
  quantum computations.
\newblock {\em Phys. Rev. Lett.}, 117:080501, Aug 2016.

\bibitem{bravyi2018quantum}
Sergey Bravyi, David Gosset, and Robert K{\"o}nig.
\newblock Quantum advantage with shallow circuits.
\newblock {\em Science}, 362(6412):308--311, 2018.

\end{thebibliography}
\newpage
\onecolumngrid
\appendix
\section{Overlap distribution}\label{supp-mat:overlap}
Here we calculate an exact expression for the overlap distribution in the case where the problem Hamiltonian is defined via the Hamming distance to the target state $\ket{t}$. The overlap is defined by 
\begin{align}
    q=\frac{1}{N}\sum_i^N s_i^{(\alpha)} s_i^{(\beta)},
\end{align}
where $\alpha$ and $\beta$ define two replicas of the system.
The probability distribution for each combination of states of the two replicas is given by the product of two Gibbs distributions at the same temperature. We calculate the probability distribution of $q$ by summing over combinations of states of the two replicas that amount to the same value of $q$,
\begin{align}
    P(q)=\sum_{i,j}^N \mathrm{e}^{-\beta H(s_i)}\mathrm{e}^{-\beta H(s_j)}\delta_{s(i)s(j),qN}.
    \label{overlapsum}
\end{align}
We note that the value $q$ is directly related to the Hamming distance, $\Delta H$, between two computational states, $q=(N-2\Delta H)/N$. For an arbitrary Hamiltonian, evaluating this sum requires exponential many classical resources as there are, in general, exponential many products of different energy pairs. However we consider a problem Hamiltonian that depends solely on the Hamming distance to the target state. Therefore we only have $N+1$ different energies and consequently only $N(N+1)/2$ different Gibbs weight pairs. In the following, we show how to calculate the overlap distribution with polynomial resources in this case.

To find all possibilities to create a certain value of $q$, we have to sum over all possibilities to create a certain distance $\Delta H$ from all possible computational states $s_i$. Let us have a look at an example, where our target state is $\ket{t}=\ket{0}^{\otimes N}$. If the system is in a computational state $l$, $\Delta_t(l)$ qubits are in the $\ket{1}$ state and consequently $N-\Delta_t(l)$ qubits are in the $\ket{0}$ state. Note that we here used $\Delta_t(l)$ as the Hamming distance from the target state w.r.t. to the state $s_i$. We now have to sum over all possible combinations, leaving us with the
\begin{align}
    P(q)=&\sum_{\Delta_t=0}^{N}\sum_{K=0}^{\Delta{H}}\binom{N}{\Delta_t}\binom{N-\Delta_t}{\Delta{H}-K}\binom{\Delta_t}{K} \mathrm{e}^{-\beta E(\Delta_t)}\mathrm{e}^{-\beta E(\Delta_t+\Delta H-2K)}.
\end{align}
This expression has only polynomial many terms and can be computed efficiently. Usually, the overlap distribution is found by doing many simulated annealing runs with different temperatures. In our case, we can set the temperature arbitrarily. We therefore have to make sure to set the temperature such that we do not hit both thermodynamic limits, where i) the ground state is occupied with certainty or ii) all states are equally likely. Therefore we shift the temperature in the regime where both cases i) and ii) do not occur. In this case, the overlap distribution gives insight into the energy landscape of the problem and the capability of QA and SA to solve the problem.
\label{appendixoverlapdist}
\end{document}